\documentclass[runningheads]{llncs}
\usepackage{graphicx}
\usepackage{amsmath,amsmath,amssymb}
\DeclareMathOperator*{\argmin}{arg\,min}
\usepackage{multirow}
\usepackage{booktabs}
\usepackage{makecell}
\usepackage{color}
\usepackage{verbatim}
\usepackage[colorlinks,
            linkcolor=red,
            anchorcolor=blue, 
            citecolor=blue,       
            ]{hyperref}
\usepackage{ragged2e}
\usepackage{marvosym}

\begin{document}
\title{Noise2SR: Learning to Denoise from Super-Resolved Single Noisy Fluorescence Image}
\titlerunning{Noise2SR: a Self-Supervised Image Denoising Method}
%
\author{Xuanyu Tian \inst{1} \and
Qing Wu \inst{1} \and
Hongjiang Wei \inst{2} \and
Yuyao Zhang \textsuperscript{1,3(\Letter)}}


\authorrunning{X. Tian et al.}
%
\institute{School of Information Science and Technology, ShanghaiTech University, \\ Shanghai, China \and
School of Biomedical Engineering, Shanghai Jiao Tong University, Shanghai, China \and
Shanghai Engineering Research Center of Intelligent Vision and Imaging, ShanghaiTech University, Shanghai, China\\
\email{zhangyy8@shanghaitech.edu.cn}
}
\maketitle              
\begin{abstract}
Fluorescence microscopy is a key driver to promote discoveries of biomedical research. However, with the limitation of microscope hardware and characteristics of the observed samples, the fluorescence microscopy images are susceptible to noise. 
Recently, a few self-supervised deep learning (DL) denoising methods have been proposed. However, the training efficiency and  denoising performance of existing methods are relatively low in real scene noise removal. To address this issue, this paper proposed self-supervised image denoising method Noise2SR (N2SR) to train a simple and effective image denoising model based on single noisy observation. 
Our Noise2SR denoising model is designed for training with paired noisy images of different dimensions. 
%
Benefiting from this training strategy, Noise2SR is more efficiently self-supervised and able to restore more image details from a single noisy observation. Experimental results of simulated noise and real microscopy noise removal show that Noise2SR outperforms two blind-spot based self-supervised deep learning image denoising methods. We envision that Noise2SR has the potential to improve more other kind of scientific imaging quality.

\keywords{Image Denoising  \and Self-Supervised Learning \and Fluorescence Microscopy Image.}
\end{abstract}

\section{Introduction}
%
Fluorescence Microscopy is  indispensable technique to boost the biomedical research in observing the spatial-temporal qualities of cells and tissues \cite{fluorescence}. It is susceptible to the influence of noise since the power of photons captured by a microscopic detector are typically weak. A direct way to promote the signal-to-noise ratio (SNR) of microscopy images is to increase the exposure time or excitation dosage. However, dynamic or real time imaging (e.g. analyzing neural activity) requires high frame rate \cite{weisenburger2019volumetric,lu2017video,skylaki2016challenges}, which limits the exposure time. While the high excitation dosage can be detrimental for sample health, and even causing inevitable damage \cite{icha2017phototoxicity,laissue2017assessing}. These contradictions can be alleviated via improvement of microscopy hardware, however, there are physical limits that are not easy to overcome. Therefore, it is of great importance to denoise and improve the quality of fluorescence microscopy images.

\par Conventional image denoising methods\cite{NLM,BM3D} require additional assumptions on noise distribution and clean image prior, which are typically general and does not leverage information of specific content of the noisy image. In recent years, a bunch of  deep learning (DL) methods based on Convolution Neural Networks (CNN) have been proposed for image denoising and the performance outperformed conventional methods \cite{RED,DnCNN}. Some Deep learning methods have been adopted by microscopists \cite{beier2017multicut,Deep-STORM,ouyang2018deep}. Weigert et al. presented the content-aware image restoration (CARE) framework in the context of fluorescence microscopy data \cite{CARE} and achieved superior performance. However, their method requires pairs of low-SNR and High-SNR images for supervised learning. High-SNR images are difficult or even unavailable in fluorescence microscopy which poses an obstacle for conventional supervised learning denoising methods. Lehtinen et al. introduced Noise2Noise (N2N)  \cite{N2N} denoising method, proposed that based on multiple noisy observations of identical image content, the requirement of clean or high-SNR image can be overcome in certain conditions. N2N can be trained with two independent noisy images of same scene and yield results close to supervised denoising learning. Unfortunately, in many scenarios of Microscopy imaging, even the acquisition of two noisy images is still difficult. Subsequently, single image self-supervised image denoising methods such as Noise2Void(N2V) \cite{N2V} and Noise2Self (N2S) \cite{N2S} have been proposed. The blind-spot network takes an image masking the center pixel as input and the value of center pixel is used as training target. But these training strategies are not efficient enough since only few pixels can contribute to the loss function\cite{laine2019high}. And the denoising performance of self-supervised methods is relatively low in real scene noise removal.\\
\indent Inspired by the recent single-image denoising works \cite{N2S,N2V,NB2NB}, we propose an effective self-supervised image denoising method Noise2SR (N2SR). 
Benefiting from the superior performance of recent image SR algorithm, our method is able to build up larger blind regions for training the denoising network, thus further improving single-image denoising performance.
Our approach consists of a sub-sampler module that generates sub-sampled noisy images from the original one; and an image SR module that improves the sub-sampled noisy image resolution to that of the original one. 
Therefore, the resolution improved sub-sampled noisy image and the rest of the original image make up the noisy image pair for self-supervised denoising network training. 
The training loss function is designed without any regularization term and to tackle the size difference issue between the paired images. 
After model training, the original noisy image goes through the denoising SR network cascading with a down-sampling operation to generate the final denoised image.  
To evaluate the performance of proposed Noise2SR, we conduct extensive experiments on publicly available Fluorescence Microscopy Denoising (FMD) dataset \cite{poissongaussiandataset} for synthetic noise and real Poisson-Gaussian noise removal. The experimental results indicate that Noise2SR outperformances blind-spot network based self-supervised image denoising methods (N2V \& N2S), and preserves more original image details from the corrupted images.

\section{Proposed Method}

\subsection{Problem Formulation}
\par Noise2Noise (N2N) is a deep-learning-based denoising model that is trained on two independent noisy observations $\{\mathbf{y}_1, \mathbf{y}_2\}$ of the same object $\mathbf{x}$, where $\mathbf{y}_1=\mathbf{x}+\mathbf{n}_1$, $\mathbf{y}_2=\mathbf{x}+\mathbf{n}_2$, and the noises $\mathbf{n}_1$ and $\mathbf{n}_2$ are i.i.d. N2N attempts to minimize the loss term of $\theta$:
\begin{equation}
\label{eq1}
    \argmin_{\theta} \mathbb{E}_{\mathbf{y}_1,\mathbf{y}_2}\lVert f_{\theta}(\mathbf{y}_1) - \mathbf{y}_2 \lVert_2^2
\end{equation}
where $f_\theta$ is the denoising network parameterized by $\theta$. N2N proves that when the expectation of noisy images ${\mathbf{y}_1,\mathbf{y}_2}$ is equal to the clean image $\mathbf{x}$, minimizing the loss term of Equ.\ref{eq1}  can converge to the same solution of supervised training.

\begin{figure}[t]
\centering
\includegraphics[width=0.95\textwidth]{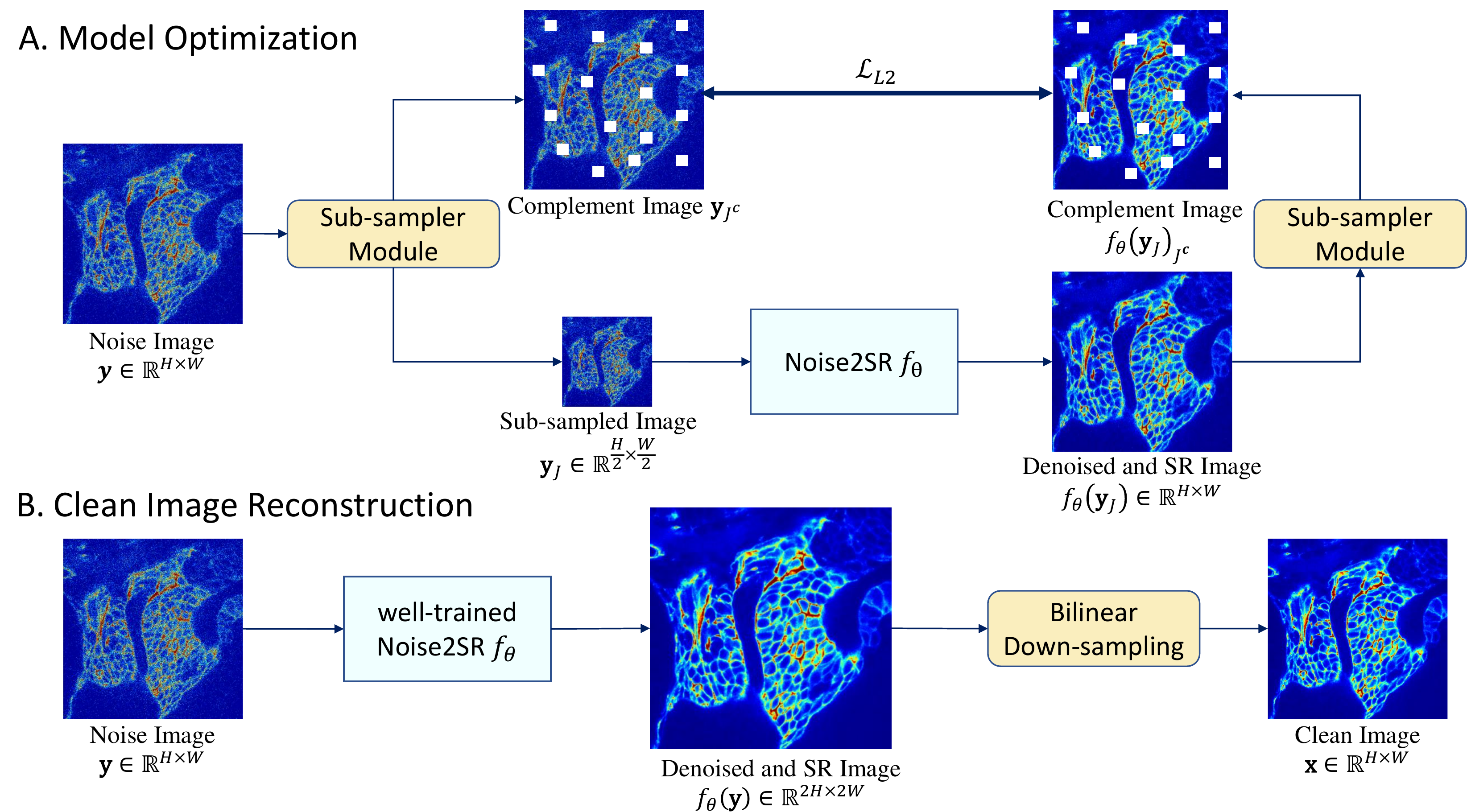}
\caption{Overview of our proposed Noise2SR model. \textbf{A. Model Optimization}: N2SR takes sub-sampled noise image $\mathbf{y}_J$ derived from sub-sampler module as input to generate the denoised and SR image $f_\theta(\mathbf{y}_J)$. Then, N2SR can be optimized by minimizing the loss computed between the complement image $\mathbf{y}_{J^c}$ and $f_\theta(\mathbf{y}_J)_{J^c}$. \textbf{B. Clean Image Reconstruction}: The well-trained N2SR takes the whole noise image $\mathbf{y}$ as input and the clean image $\mathbf{x}$ can be generated from the denoised and SR image $f_\theta(\mathbf{y})$ with down-sampling operation. } 
\label{fig1}
\end{figure}

\noindent \textbf{Paired Noisy Images with Different Resolution.} Pipeline of the proposed method is demonstrated in Fig.\ref{fig1}. N2N training requires at least two independent noisy observations of a same object which limits its application scenario. Thus, we propose a sub-sampler module (Sec.\ref{Sub-sampler-module}) to extract sub-sample image $\mathbf{y}_J$ from the noisy image $\mathbf{y}$. Our Noise2SR model is trained with paired noisy images with different resolution to learn image denoising.
With different dimensions of $\mathbf{y}_J$ and $\mathbf{y}$, the model can not learn an identity mapping. Thus an additional image SR module with up-sampling scale 2 is composed into the denoising network $f_\theta$.
Specifically, the Noise2SR network takes sub-sample image $\mathbf{y}_J$ as input and the complement part $\mathbf{y}_{J^c}$ in image $\mathbf{y}$ as labeling for training (Sec.\ref{Optimization}). We thus extend the loss function term Equ.\ref{eq1} to $\argmin_\theta \mathbb{E}_{\mathbf{x},\mathbf{y}} \lVert f_\theta(\mathbf{y}_J)_{J^c} - \mathbf{y}_{J^c} \lVert_2^2$, where ${J^c}$ is the complement region of the sub-sampler region ${J}$ in the noisy image $\mathbf{y}$.
\begin{theorem}
\label{theorem1}
Let $\mathbf{y} = \mathbf{x} + \mathbf{n}$ be an image corrupted by zero-mean noise with variance $\sigma_\mathbf{n}^2$. $\mathbf{y}_{J}$ is subset of the noisy image $\mathbf{y}$ and $\mathbf{y}_{J^c}$ is the complement of $\mathbf{y}_{J}$. Suppose the noise $\mathbf{n}$ is independent to clean image $\mathbf{x}$ and the noise $\mathbf{n}$ is pixel-wise independent. Then it holds that
\begin{equation}
    \begin{split}
    \mathbb{E}_{\mathbf{x},\mathbf{y}} \lVert f_\theta(\mathbf{y}_J)_{J^c} - \mathbf{y}_{J^c} \lVert_2^2
    & = \mathbb{E}_{\mathbf{x},\mathbf{y}} \lVert f_\theta(\mathbf{y}_J)_{J^c} - \mathbf{x}_{J^c} \lVert_2^2 + \sigma_\mathbf{n}^2 
   \end{split}
\end{equation}
\end{theorem}
Theorem \ref{theorem1} states that the optimizing $\argmin_{\theta} \mathbb{E}_{\mathbf{x},\mathbf{y}} \lVert f_\theta(\mathbf{y}_J)_{J^c} - \mathbf{y}_{J^c} \lVert_2^2$ yields the same solution as the supervised training (The proof is given in the supplementary material). On the basis of this conclusion, we propose a self-supervised denoising method Noise2SR.


\subsection{Sub-sampler module}
\label{Sub-sampler-module}
\begin{figure}[h]
\centering
\includegraphics[width=0.75\textwidth]{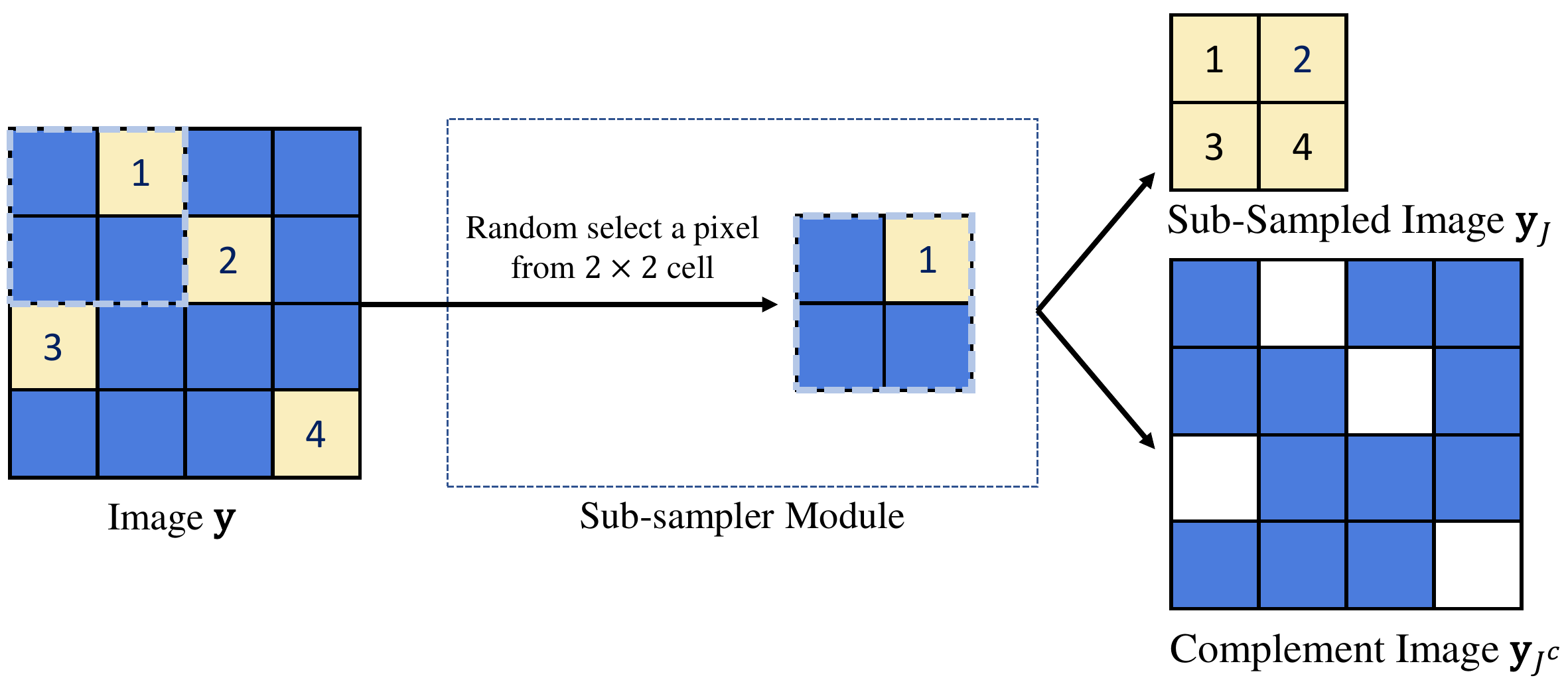}
\caption{Workflow of the sub-sampler module for a noise input image example of $4\times4$.} \label{fig2}
\end{figure}

The goal of sub-sampler module is to generate the sub-sampled image $\mathbf{y}_J$ and complement image $\mathbf{y}_{J^c}$ from the noisy input image $\mathbf{y}$. Fig.\ref{fig2} illustrates the workflow of the sub-sampler module for a noisy input example of $4\times4$. 
The details of sub-sampler are described below:
\begin{itemize}
    \item [1.]
    The image $y \in \mathbb{R}^{H\times W}$ is divided into a number of $\lfloor H/2 \times W/2 \rfloor$ cells with each cell of size $2\times 2$.
    \item [2.]
    The $(i,j)$-th pixel of sub-sampled image $\mathbf{y}_J \in \mathbb{R}^{\lfloor H/2 \times W/2 \rfloor}$ is randomly selected from the $i$-th row and $j$-th column cell. 
    \item [3.]
    The complement image $\mathbf{y}_{J^c}$ masks the selected pixels from $\mathbf{y}$ which can be expressed as $\mathbf{y}_{J^c}=\mathbf{m}_J\odot y$, where $\mathbf{m}_J$ is binary mask and $\odot$ is Hadamard product.
\end{itemize}

\subsection{Model Optimization}
\label{Optimization}
The pipeline of model optimization is shown in the Fig.\ref{fig1} \textbf{A}. The sub-sampled noisy image $\mathbf{y}_J \in \mathbb{R}^{H/2\times W/2}$ and the complement image $ \mathbf{y}_{J^c} \in \mathbb{R}^{H\times W} $ (positions of white spots are filled with zeros) are firstly extracted from the original noisy input image $\mathbf{y} \in \mathbb{R}^{H\times W}$ through the sub-sampler module. 
Then, the SR denoising network is optimized by minimizing the prediction error $\mathcal{L}_{\mathrm{L}2}$ between the complement image $\mathbf{y}_{J^c}$ and the output of N2SR-net $f_\theta(\mathbf{y}_J)_{J^c}$. The loss function $\mathcal{L}_{\mathrm{L}2}$ is denoted as below:
\begin{equation}
    \mathcal{L}_{\mathrm{L}2} = \frac{1}{m} \sum_{i=1}^{m}(f_\theta(\mathbf{y}_J)_{J^c} - \mathbf{y}_{J^c})^2
\end{equation}
where $m$ is the mini-batch size during each training iteration.

\noindent \textbf{Network Architecture}
Following the previous works \cite{N2N,N2S,N2V}, we employ U-Net \cite{U-net} as the backbone network of our proposed N2SR model.
To generate the denoised SR image with same dimensions as the original noisy image $\mathbf{y}$ from the noisy input sub-sampled image $\mathbf{y}_J$, we propose a simple SR module including a transposed convolutional layer of $2\times$ factor \cite{conv} and two convolutional layers of $1\times 1$ kernel size after the U-Net. Specially, we set the output channel of U-Net as 256 allowing the network to propagate context information to the transpose convolution layer. The transpose convolutional layer is used to implement the upsampling operation and the last two convolutional layers are used to generate the denoised SR result.

\subsection{Clean Image Reconstruction}
The pipeline of clean image reconstruction using well-trained N2SR model is shown in Fig.\ref{fig1}.\textbf{B}, which is presented into two steps: ({\romannumeral1}). We first feed the noise input image $\mathbf{y} \in \mathbb{R}^{H\times W}$ into the N2SR network to generate the denoised and super-resolution result $f_\theta (\mathbf{y}) \in \mathbb{R}^{2H\times 2W}$;  ({\romannumeral2}). We then apply bilinear down-sampling operator on image $f_\theta(\mathbf{y})$ to get the desired clean image $\mathbf{x} \in \mathbb{R}^{H\times W}$ with same dimensions as the input noisy image $\mathbf{y}$. In supplementary material, we provide the comparisons of the denoised and super-resolution results by our model and other denoising methods combined with interpolation methods.

\section{Experiment}
\subsection{Setup}
\noindent \textbf{Datasets.}
We evaluate the proposed Noise2SR on both simulated noise and real noise removal experiments using the Fluorescence Microscopy Denoising (FMD) dataset  \cite{poissongaussiandataset}. In FMD dataset, each sample is scanned 50 times to generate 50 noisy observations. The averaging image of the 50 images is then used as high-SNR reference image. In \textbf{simulated noise evaluation} experiment, we add simulated additive Gaussian noise $(\sigma = 15)$, Poisson noise $(\lambda = 20)$ and Poisson-Gaussian noise $(\sigma = 5, \lambda = 15)$ in the averaging high-SNR images respectively, and the averaging images are considered as denoising ground truth (GT) in this case. For each simulated noise dataset, we generate two indepentent noisy images for 180 samples and split them into three parts: 270 images in training set, 54 images for validation set and 36 images for test set. 
In the \textbf{real scene noise evaluation} experiment, we use the confocal and two-photon microscopy data with noise level 1 (raw noisy image without any processing) in FMD dataset, which consisting 9000 images in 9 imaging configurations (combination of microscopy modalities and biological samples). For each imaging configuration, we randomly selected one noisy image from \#1 to \#18 FOV and split them into training and validation set. The test set consists of 3 images respectively selected from \#19 and \#20 FOV in each imaging configuration. Thus, there are 54 images in the test set.

\noindent \textbf{Baseline \& Metrics.}
We employ one classic unsupervised denoiser BM3D \cite{BM3D1,BM3D2} and three deep learning denoisers including Noise2Noise (N2N) \cite{N2N}, Noise2Void (N2V) \cite{N2V} and Noise2Self (N2S) \cite{N2S} as baselines. \textit{Note that N2N is trained on independent noisy image pairs while N2V, N2S and the proposed N2SR are trained on a single noisy image.} The three deep learning models are implemented by PyTorch framework \cite{pytorch} and the network structure is U-Net following the same structure as Noise2Noise \cite{N2N}. Considering the constraint of GPU Memory and training efficiency, we randomly crop  $128\times 128$ image patches from the original images of size  $512\times 512$ for every training iteration. We evaluate the performance of denoising results quantitatively using two metrics: Peak Signal-to-Noise (PSNR) and Structural Similarity Index Measure (SSIM).

\subsection{Results}

\begin{figure}[t]
\includegraphics[width=\textwidth]{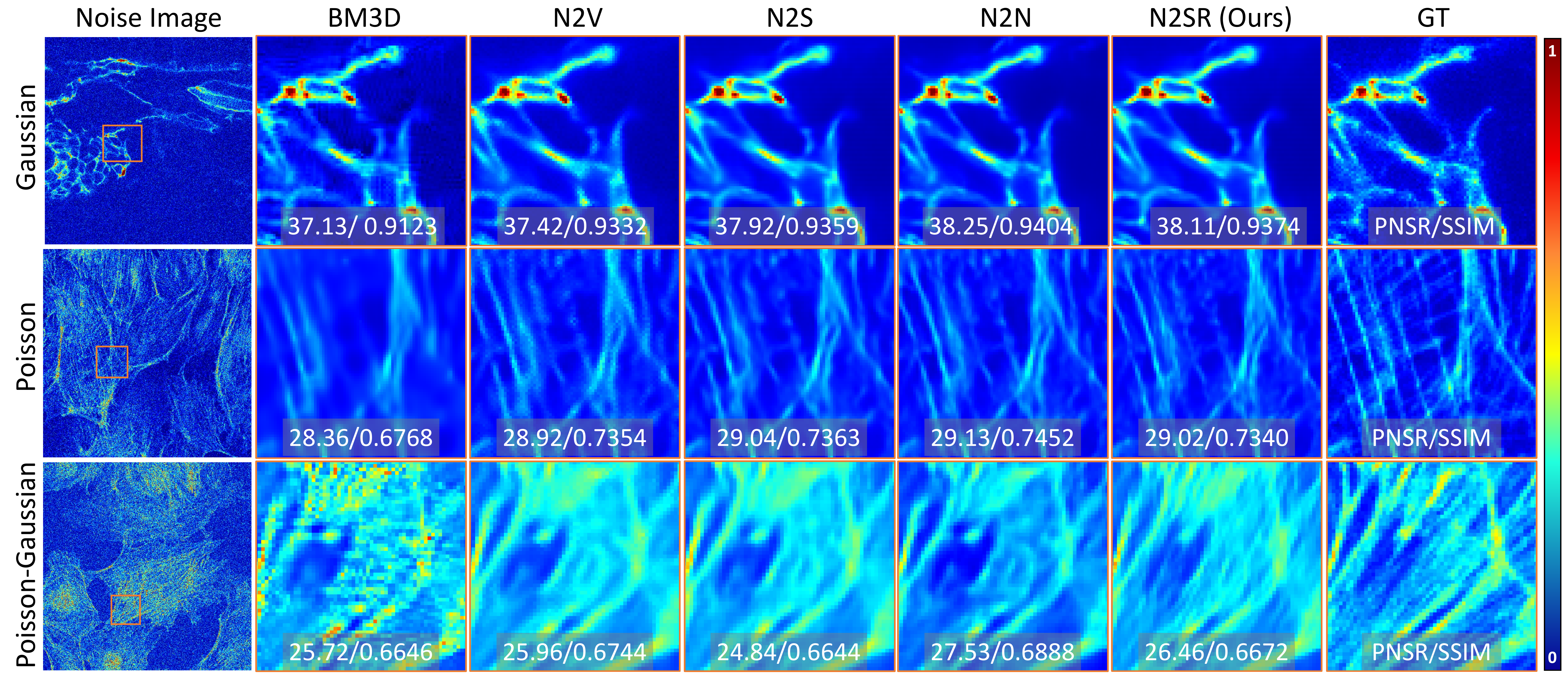}
\caption{Qualitative result of simulated noise experiment with all the compared methods on confocal microscopy data of zebrafish embryo cells with additive Gaussian noise $(\sigma = 15)$ and BPAE(F-actin) data with Poisson $(\lambda = 20)$ and  Poisson-Gaussian  $(\sigma = 5, \lambda = 15)$ noise. } 
\label{fig3}
\end{figure}

\begin{table}[b]
\centering
\caption{Performance of simulated noise experiment with  all compared method denoising results (PSNR (dB) /SSIM) on the test set. The best and second performance are highlighted in \textcolor[RGB]{255,0,0}{red} and \textcolor[RGB]{0,0,255}{blue}, respectively.}
\resizebox{0.90\textwidth}{!}{
\begin{tabular}{c|c|c|c|c|c} 
\toprule
\multirow{2}{*}{Noise Type} & \multicolumn{5}{c}{Methods}                                                                                                                                       \\ 
\cline{2-6}
                            & BM3D         & N2V          & N2S                            & N2SR (Ours)                                      & N2N                                             \\ 
\hline\hline
Gaussian     $(\sigma = 15)$                & 34.15/0.8572 & 34.56/0.8871 & 34.99/\textcolor{blue}{0.8938} & \textcolor{blue}{35.27}/0.8936                   & \textcolor{red}{35.42}/\textcolor{red}{0.9002}  \\ 
\hline
Poisson $(\lambda = 20)$                     & 31.32/0.8601 & 33.27/0.8801 & 33.45/0.8812                   & \textcolor{blue}{33.50}/\textcolor{blue}{0.8827} & \textcolor{red}{33.62}/\textcolor{red}{0.8873}  \\ 
\hline
\makecell{mix Poisson $(\lambda = 15)$ \\ \& Gaussian $(\sigma = 5)$}        & 28.01/0.8271 & 31.69/0.8512 & 31.43/0.8461                   & \textcolor{blue}{31.80}/\textcolor{blue}{0.8523} & \textcolor{red}{32.56}/\textcolor{red}{0.8633}  \\
\bottomrule
\end{tabular}}
\label{tab1}
\end{table}
\noindent \textbf{Validation Study on the simulated noise experiment dataset.} In Table \ref{tab1}, we show the quantitative results of all methods on the test set of the simulation noise data. To sum up, N2N achieves the best performance and our self-supervised N2SR obtains suboptimal results in most cases.  Fig.\ref{fig3} demonstrates the qualitative evaluation results of all the compared methods. Compared with self-supervised N2V, N2S and even N2N, our N2SR preserves more image details, and the detail textures are quite consistent with that in the GT images. When there is only Gaussian noise added, all DL methods perform similar and stable. While under additive Poisson and mixed Gaussian-Poisson noise, the advantages of the proposed N2SR are more prominent in both qualitative and quantitative assessments. Theoretically, the microscopy imaging system is affected by mixed Gaussian-Poisson background noise. Based on the simulation data results we envision N2SR is more efficient than other single-image denoising methods to reduce real scene microscopy image noise.

\begin{table}[b]
\centering
\caption{Quantitative results (PSNR (dB) /SSIM) of all compared method on the test set of the FMD dataset with noise level 1 (raw).}
\resizebox{0.85\textwidth}{!}{
\begin{tabular}{c|l|c|c|c|c|c} 
\toprule
\multirow{2}{*}{Microscopy} & \multicolumn{1}{c|}{\multirow{2}{*}{Samples}} & \multicolumn{5}{c}{Methods}                                                                                                                      \\ 
\cline{3-7}
                            & \multicolumn{1}{c|}{}                         & BM3D         & N2V          & N2S          & N2SR(Ours)                                       & N2N                                              \\ 
\hline\hline
\multirow{5}{*}{Confocal}   & BPAE (Nuclei)                                 & 34.63/0.9683 & 39.04/0.9784 & 39.42/0.9800 & \textcolor{red}{39.69}/\textcolor{blue}{0.9822}  & \textcolor{blue}{39.67}/\textcolor{red}{0.9828}  \\
                            & BPAE(F-actin)                                 & 32.97/0.8956 & 33.43/0.8917 & 34.02/0.9082 & \textcolor{red}{34.45}/\textcolor{blue}{0.9114}  & \textcolor{blue}{34.44}/\textcolor{red}{0.9150}  \\
                            & BPAE (Mito)                                   & 31.16/0.9301 & 32.27/0.9303 & 33.06/0.9405 & \textcolor{blue}{33.47}/\textcolor{blue}{0.9443} & \textcolor{blue}{33.61}/\textcolor{red}{0.9484}  \\
                            & Zebrafish                                     & 29.09/0.8732 & 33.97/0.9098 & 34.06/0.9108 & \textcolor{red}{34.14}/\textcolor{blue}{0.9131}  & \textcolor{red}{34.14}/\textcolor{red}{0.9167}   \\
                            & Mouse Brain                                   & 33.07/0.9457 & 37.11/0.9603 & 37.40/0.9612 & \textcolor{blue}{37.49}/\textcolor{blue}{0.9651} & \textcolor{red}{37.74}/\textcolor{red}{0.9652}   \\ 
\hline\hline
\multirow{4}{*}{Two-Photon} & BPAE (Nuclei)                                 & 29.12/0.9120  & 32.91/0.9202 & 32.92/0.9216 & \textcolor{blue}{33.17}/\textcolor{red}{0.9334}  & \textcolor{red}{33.23}/\textcolor{blue}{0.9328}  \\
                            & BPAE (F-actin)                                & 27.98/0.7653 & 28.36/0.7678 & 28.48/0.7833 & \textcolor{blue}{28.66}/\textcolor{blue}{0.7916} & \textcolor{red}{28.93}/\textcolor{red}{0.7970}   \\
                            & BPAE (Mito)                                   & 29.79/0.8596 & 31.42/0.8622 & 31.33/0.8660 & \textcolor{blue}{31.72}/\textcolor{blue}{0.8754} & \textcolor{red}{31.89}/\textcolor{red}{0.8817}   \\
                            & Mouse Brain                                   & 31.38/0.9088 & 33.69/0.9046 & 33.82/0.9059 & \textcolor{blue}{34.25}/\textcolor{red}{0.9194}  & \textcolor{red}{34.27}/\textcolor{blue}{0.9185}  \\
\bottomrule
\end{tabular}}
\label{tab2}
\end{table}

\begin{figure}[ht]
\centering
\includegraphics[width=0.95\textwidth]{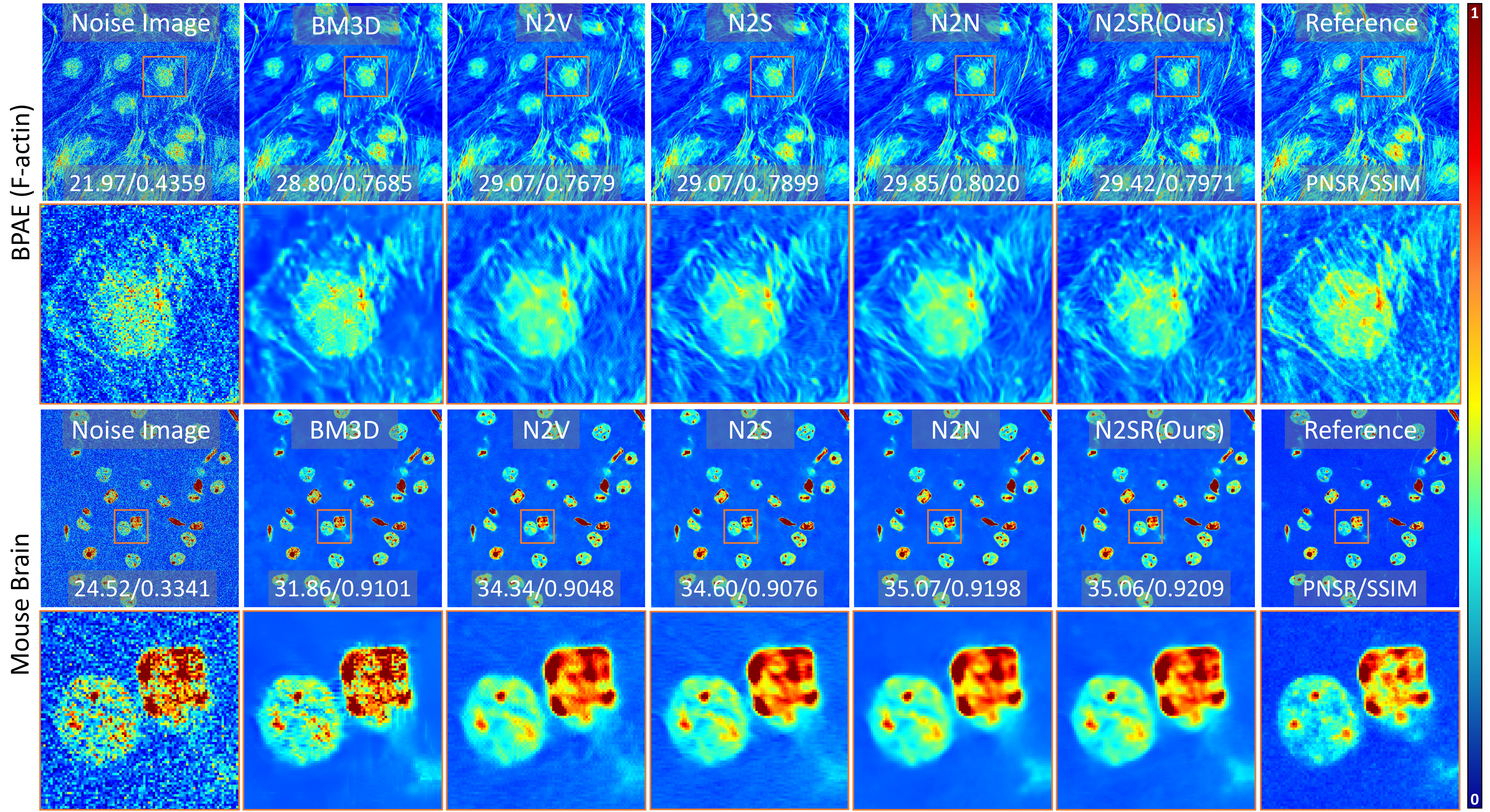}
\caption{Qualitative results of all the compared methods on two-photon microscopy data (BPAE (F-actin) and Mouse Brain).} \label{fig4}
\end{figure}

\noindent \textbf{Performance on the Real scene experiment dataset.} In Table \ref{tab2}, we demonstrate the quantitative comparison of all the compared methods on the test set of real scene experiment dataset. The high-SNR reference image is considered as ground truth for computing the evaluation metrics. Overall, our N2SR significantly outperforms the other two self-supervised N2V and N2S networks and the traditional denoiser BM3D. Moreover, compared with N2N, N2SR achieves a comparable performance. Fig.\ref{fig4} illustrates the Qualitative results of all the compared methods on two representative test samples (BPAE (F-actin), \#19 FOV, and Mouse Brain, \#19 FOV). On the BPAE (F-actin) test sample, our N2SR preserves the sharpest image details which are very similar to those in the reference image.  On the test sample Mouse Brain, we observe that only our N2SR and N2N produce clean image background while well maintaining clear foreground tissues and sharp tissue boundaries. In contrast, N2V and N2S cannot completely remove background noise; N2V even produced fuzzy artifacts on the tissue boundary. As the microscopy imaging system produces mixed Gaussian-Poisson background noise, the well-behaved real scenes denoised result is consistent with what we found in the simulation experiment.

\section{Conclusion}
In this paper, we propose Noise2SR, a self-supervised denoising method for single noisy fluorescence image denoising task. Our method does not require multiple noisy observations and external noise distribution assumptions. We propose a sub-sampler and an SR module in the denoising network, assuring that Noise2SR can be trained with constructed paired noisy images with different image dimensions. Our method builds up a larger blind region for training the denoising network compared with other existing blind-spot networks, thus further improving the training efficiency and single-image denoising task performance. The qualitative and quantitative results from simulated noise and real Poisson-Gaussian noise removal of FMD dataset show that out Noise2SR outperforms a conventional denoiser BM3D and two blind-spot based self-supervised image denoising methods and achieves comparable performance to Noise2Noise.

\subsubsection{Acknowledgements.} This study is supported by the National Natural Science Foundation of China (No. 62071299, 61901256, 91949120).

%
%
%
\bibliographystyle{splncs04}
\bibliography{ref}


\end{document}